\documentclass[onecolumn,prl,floatfix,amssymb,aps,superscriptaddress,showpacs]{revtex4}
\usepackage{graphicx}
\usepackage{color}
\usepackage{epsfig}
\usepackage{amsmath}
\usepackage{amssymb}
\usepackage{mathrsfs}
\usepackage{bm}
\usepackage{subfigure}
\usepackage{float}
\usepackage{url}
\usepackage{enumerate}
\usepackage{braket}
\usepackage{comment}
\usepackage{bibunits}

\definecolor{violet}{rgb}{0.56,0.0,1.0}

\begin{document}

\begin{bibunit}

\hsize\textwidth\columnwidth\hsize\csname@twocolumnfalse\endcsname

\title{Symmetry-protected topological phase in a one-dimensional correlated bosonic model with a synthetic spin-orbit coupling}

\author{Jize Zhao}
\affiliation{Institute of Applied Physics and Computational Mathematics, Beijing 100088, China}

\author{Shijie Hu}
\affiliation{Department of Physics and Research Center Optimas, Technical University Kaiserslautern, 67663 Kaiserslautern, Germany}

\author{Ping Zhang}
\affiliation{Institute of Applied Physics and Computational Mathematics, Beijing 100088, China}

\begin{abstract}
By performing large-scale density-matrix renormalization group simulations, we investigate 
a one-dimensional correlated bosonic lattice model with a synthetic spin-orbit coupling
realized in recent experiments. In the insulating regime, this model exhibits a symmetry-protected topological phase. 
This symmetry-protected topological phase is stabilized by time-reversal symmetry 
and it is identified as a Haldane phase. We confirm our conclusions further by analyzing the entanglement spectrum.
In addition, we find four conventional phases: a Mott insulating phase with no long range order, 
a ferromagnetic superfluid phase, a ferromagnetic insulating phase and a density-wave phase.
\end{abstract}

\maketitle

In recent years, symmetry-protected topological (SPT) phases \cite{GU1, POLLMANN1, SCHUCH1, CHEN1, CHEN2} have been actively 
studied in condensed matter physics. One of the most famous SPT phase may be the topological insulator \cite{HASAN1, QI1},
where the phase is stabilized by $U(1)$ symmetry as well as time-reversal symmetry. 
Understanding such exotic state of matter usually involves the spin-orbit coupling (SOC). 
A well-tuned SOC may change drastically the Fermi surface as well as the 
dispersion of the energy band, leading to exotic low-energy phenomena \cite{KANE1}. 
However, in materials, the SOC is usually quite small and manipulating the SOC is nontrivial. 
Recently, by using one pair of tunable lasers, a highly manipulatable SOC with equal Rashba and Dresselhaus weight was realized 
in both ultracold boson and fermion atoms \cite{LIN1,WANG1,CHEUK1,ZHANG1}. These pioneering experiments triggered growing interest 
in this area \cite{WU1,BIJL1,ISKIN1,ZHAI1,COLE1,HU1,HAN1}, providing more promising platforms to 
study exciting physical phenomena in ultracold atoms with an SOC \cite{JOTZU1, BEELER1, JI1, FU1, JIMENEZ1, LUO1}. 

The SOC realized in Bose-Einstein condensate is written as $\sim p_x\sigma_y$. Subject to an optical lattice, 
its tight-binding form is given by   
\begin{equation}
\mathcal{T}_\text{soc} = -\lambda\sum_i(\hat{c}^{\dagger}_{i\uparrow}\hat{c}_{i+1\downarrow} -\hat{c}^{\dagger}_{i\downarrow}\hat{c}_{i+1\uparrow})+h.c.,
\end{equation}
where $\hat{c}_{i\tau}$ is the boson annihilation operator at site $i$ with spin $\tau$, $\lambda$ is the 
SOC strength. Together with the kinetic energy $\mathcal{K} = -t\sum_{i\tau} \left(\hat c_{i\tau}^\dagger \hat c_{i+1\tau}+h.c.\right)$, 
it is known that in this case the SOC can be eliminated by a local gauge transformation \cite{CAI1}, resulting in a renormalized  
hopping term $t\rightarrow\sqrt{t^2+\lambda^2}$. 
However, the low-energy dynamics of bosons in optical lattice is usually captured by correlated lattice models. 
The interaction is in general not invariant under such a local gauge transformation \cite{ZHAO1, ZHAO2, SUPP1}. 
The interplay among the kinetic energy, the SOC and the onsite interaction can lead to 
a variety of quantum phases \cite{HO1, ZHAO1,ZHAO2,PIRAUD1,XU1,PEOTTA1}.

On the other hand, in fermionic systems, SOC is known to play an important role in many topologically nontrivial phases
and has been extensively studied in the past decade.
In contrast, as far as we know, the related phenomena in bosonic systems with an SOC remain unexplored yet. 
In addition, it has been well established that there is a topologically nontrivial phase
in the one-dimensional extended Bose-Hubbard model \cite{TORRE1,BERG1,ROSSINI1,BATROUNI1,EJIMA1}. 
This phase is stabilized by inversion symmetry \cite{TORRE1,EJIMA1}. 
In the presence of an SOC, inversion symmetry is explicitly broken. 
In such a situation, whether a topologically nontrivial phase can survive is unclear so far.
Very recently, the spin-orbit coupled Bose-Einstein condensate was loaded into one-dimensional optical lattices \cite{HAMNER1}.
These motivate us to investigate a one-dimensional correlated bosonic model with an SOC to search for a possible bosonic SPT phase in
analogy to the topological insulator. 
Through extensive numerical analysis,  we found that such a topologically 
nontrivial phase indeed exists. We consider a general one-dimensional model with the SOC realized in experiments \cite{LIN1}, given by 
\begin{eqnarray}
\mathcal{H} & = & \mathcal{K} + \mathcal{T}_\text{soc}
                  +\frac{U}{2}\sum_{i\tau} \hat n_{i\tau}( \hat n_{i\tau}-1) \nonumber \\
            &   & + U^{\prime} \sum_i \hat n_{i\uparrow} \hat n_{i\downarrow} + V\sum_{i\tau} \hat{n}_{i\tau}\hat{n}_{i+1\tau} \nonumber \\
            &   & +V^\prime\sum_{i\tau} \hat{n}_{i\tau} \hat{n}_{i+1\bar{\tau}} -\mu\sum_{i}\hat n_{i},
\label{HSOC}
\end{eqnarray}
where $\mathcal{K}$ and $\mathcal{T}_\text{soc}$ are the kinetic energy and the SOC, respectively, as we mentioned previously. 
$i$ runs from 1 to $L$ with $L$ the chain length.
$\hat{n}_{i\tau}=\hat{c}^\dagger_{i\tau}\hat{c}_{i\tau}$ is the boson number operator with spin $\tau$ at site $i$.
$\bar{\tau}$ is the opposite spin of $\tau$. $\hat{n}_{i}=\hat{n}_{i\uparrow}+\hat{n}_{i\downarrow}$ 
is the total particle number operator at site $i$.
$U$ and $U^\prime$ are the on-site intracomponent and intercomponent interaction, respectively.
$V$ and $V^\prime$ are the nearest-neighbor intracomponent and intercomponent interaction, respectively. 
The nearest-neighbor interaction may be realized by the dipole-dipole interaction \cite{DENG1, WILSON1, NG1, SYZRANOV1}. 
$\mu$ is the chemical potential, controlling the filling factor. 
 
In addition to the $U(1)$ symmetry corresponding to the conservation of total particle number, 
the Hamiltonian (\ref{HSOC}) has time-reversal symmetry, invariant under the transformation
$\mathcal{T}c_{i\tau}\mathcal{T}^{-1}=\left(i\sigma_y\right)_{\tau\tau^\prime}c_{i\tau^\prime}$.
Moreover, one can interchange $t$ and $\lambda$ as well as $V$ and $V^\prime$ by the following transformation \cite{ZHAO1}
\begin{eqnarray}
&&\hat c_{i \tau}~~~\rightarrow ~~{\rm sign}_\tau ~\hat c_{i \tau}, \ \  \ \ ~\hat c_{i+1 \tau} \rightarrow ~~\hat c_{i+1 \bar{\tau}}, \nonumber  \\
&&\hat c_{i+2 \tau} \rightarrow -{\rm sign}_\tau ~\hat c_{i+2 \tau}, \ \ \hat c_{i+3 \tau} \rightarrow -\hat c_{i+3 \bar{\tau}} 
\label{EXCHANGE}
\end{eqnarray}
for every 4-sites with ${\rm sign}_\uparrow=1$ and ${\rm sign}_\downarrow=-1$. 
Thanks to this transformation, we only need to consider the parameter regime satisfying $t/\lambda\le{1}$
while phases for those $t/\lambda>{1}$ are immediately available. 

\begin{figure}[h]
\includegraphics[width=7.7cm, clip]{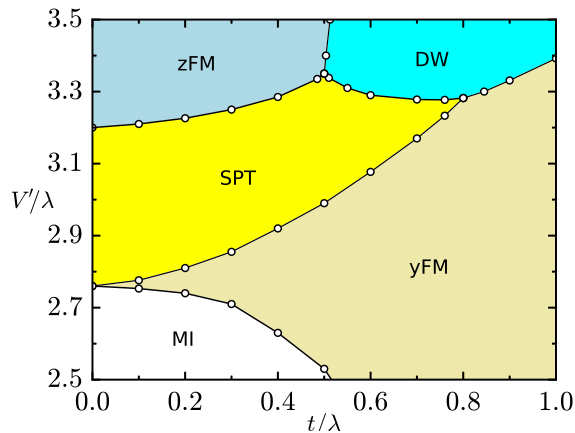}
\vspace{-2.5cm}
\caption{(color online) A typical ground-state phase diagram
is shown in $(t/\lambda, V^\prime/\lambda)$ plane for the given constraint (see text).
The other parameters are $U/\lambda=5$, $U^\prime/\lambda=0.1$ and $V/\lambda=0.02$.
This phase diagram includes a Mott insulating (MI) phase, a ferromagnetic superfluid phase polarized in y direction (yFM),
a symmetry-protected topological phase (SPT), a ferromagnetism polarized in z direction (zFM),
a density wave (DW) phase.}
\label{FIG1}
\end{figure}
In general, the Hamiltonian (\ref{HSOC}) is not exactly solvable. Thus, we resort to
density-matrix renormalization group (DMRG) \cite{WHITE1, PESCHEL1, SCHOLLWOCK1, MCCULLOCH1} to study it numerically.  
In our work, we restrict the average particle density $\rho$ to $\rho=2$, i.e. two particles per site.   
The maximum degree of freedom at each site is truncated to 9. 
We also verified that larger truncation does not alter our results qualitatively in the deep insulating regime.
The open boundary condition (OBC) is used unless stated explicitly otherwise.
In most simulations, the states we kept range from 800 to 1500, depending on the parameters and quantities we are interested in.  
For better numerical accuracy, we will focus our study on the regime with $\lambda,t$ much smaller than $U$.

A typical phase diagram is shown in Fig. \ref{FIG1} in the $(t/\lambda, V^\prime/\lambda)$ plane. 
We found five phases in the phase diagram. Three of these phases are long-range ordered. 
They are marked as yFM, zFM, and DW. yFM is a ferromagnetic phase polarized in the y direction. 
zFM is a ferromagnetic phase polarized in the z direction. DW is a phase with 
a long-range atomic density-wave order. The other two phases (MI and SPT) have no long-range orders.
MI represents a Mott insulating phase. SPT is a symmetry-protected topological phase, which will be finally 
identified as the Haldane phase. Of all these phases, yFM is a superfluid phase and others are insulating phases\cite{SUPP1}.
The transitions of zFM-DW and yFM-DW are of the first order, and other phase transitions are continuous.
The criticality of these continuous phase transitions can be understood from the
central charge $c$ \cite{CALABRESE1} and the symmetries. Our results are summarized below, and details of our calculations and analysis
are present in the Supplemental Material\cite{SUPP1}.
\begin{figure}[h]
\includegraphics[width=7.7cm, clip]{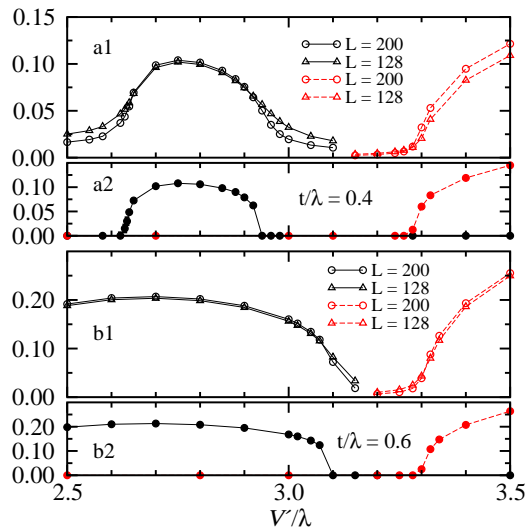}
\caption{(color online) Order parameters are shown as a function of $V^\prime/\lambda$. 
Panel a1: the order parameters $\mathcal{S}^y\left(0\right)$ (symbols connected by black solid lines) 
and $\mathcal{S}^z\left(0\right)$ (symbols connected by red dashed lines) for L = 128 and L = 200
are shown for $t/\lambda = 0.4$. Panel a2: the corresponding order parameters are extrapolated to the thermodynamic limit
with L = 96, 128 , 160, 200 (in the vicinity of the transition points, L = 320 and 400 are also used).
The three transition points are determined to be $V^\prime/\lambda=2.63(1), 2.93(1)$, and $3.27(2)$.
Panel b1 and b2: similar to panels a1 and a2, but the order parameters 
are $\mathcal{S}^y\left(0\right)$(symbols connected by black solid lines) 
and $\mathcal{D}\left(\pi\right)$(symbols connected by red dashed lines) for $t/\lambda=0.6$. 
The two transition points are $V^\prime/\lambda=3.08(2)$ and $3.29(2)$. In panels a1 and a2, $\mathcal{S}^z(0)$
is divided by 4. In panels b1 and b2, $\mathcal{D}(\pi)$ is divided by 8.}
\label{FIG2}
\end{figure}
$c$ at the SPT-zFM and the SPT-DW transition points are 1/2, indicating an Ising universality class.
At $t/\lambda=0$, $c=2$ at the MI-SPT transition point. Its low-energy behavior is described by a two-component Tomonaga-Luttinger liquid.
In the yFM phase, $c=1$, suggesting a one-component Tomonaga-Luttinger liquid. At both MI-yFM and yFM-SPT transition points,
$c$ is 1. These transitions are expected to be the XY type. A similar behavior has also been predicted in
another two-component model \cite{EJIMA2}. Below we will discuss the phase diagram further. For simplicity, we
restrict our discussions to the parameter regime in Fig. \ref{FIG1}.

The phases in Fig. \ref{FIG1} with long-range orders can be conveniently identified by corresponding local order parameters. 
For this purpose, we will study the spin-spin and density-density correlation functions as well as 
their Fourier transformation, defined by 
\begin{eqnarray}
\mathcal{S}^\nu\left(q\right)=\frac{1}{L^2}\sum_{ij}e^{iq\left(i-j\right)}\langle\hat S^{\nu}_{i} \hat S^{\nu}_{j}\rangle,\\\
\mathcal{D}\left(q\right)=\frac{1}{L^2}\sum_{ij}e^{iq\left(i-j\right)}\langle\left(\hat{n}_{i}-\rho\right)\left(\hat{n}_{j}-\rho\right)\rangle,
\end{eqnarray}
with $\hat{S}_{i}^{\nu}= \sum_{\tau\tau^\prime}\hat{c}_{i\tau}^\dagger\sigma^{\nu}_{\tau\tau^\prime}\hat{c}_{i\tau^\prime}/2$, 
where $\sigma^\nu$ is the Pauli matrix and $\nu=y$ or $z$. The order parameters in yFM, zFM and DW phases are given by
$\mathcal{S}^y(0)$, $\mathcal{S}^z(0)$ and $\mathcal{D}(\pi)$, respectively. In Fig. \ref{FIG2}, we plot the order parameters 
as a function of $V^\prime/\lambda$. The phase transition is signaled  by the vanishing of the order parameters in
the thermodynamic limit. In panels a1 and a2, $\mathcal{S}^y(0)$ and $\mathcal{S}^z(0)$ are shown for $t/\lambda = 0.4$. 
We can conclude that for $2.63<V^\prime/\lambda<2.93$, it is in a yFM phase, and for $V^\prime/\lambda>3.27$ it is 
in a zFM phase. In panels b1 and b2, we plot the order parameters $\mathcal{S}^y(0)$ and $\mathcal{D}(\pi)$ for $t/\lambda=0.6$ as 
a function of $V^\prime/\lambda$. We can then determine that for $V^\prime/\lambda<3.08$ it is in a yFM phase 
and for $V^\prime/\lambda>3.29$ it is in a DW phase.

To gain a deeper understanding of the phase diagram in Fig. \ref{FIG1}, in particular, the SPT phase,
we calculate the neutral excitation gaps. They are defined as $\Delta_k = E_k(N,L)-E_0(N,L)$,
where $E_k(N,L)$ and $E_0(N,L)$ are the energy of the $k$-th excited state and the ground state, respectively,
for boson number $N$ and length $L$. In our work, $N$ is fixed to be $2L$.
\begin{figure}[h]
\includegraphics[width=7.7cm, clip]{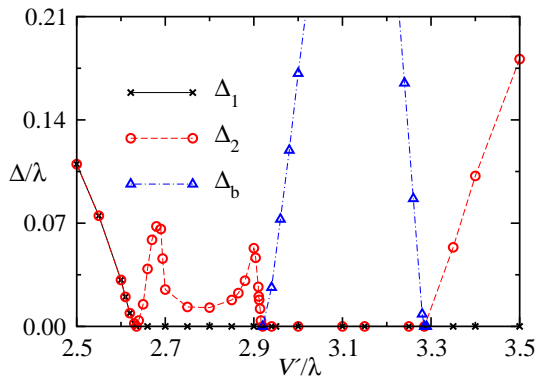}
\caption{(color online) Excitation gaps for $t/\lambda=0.4$ are shown as a function of $V^\prime$. 
The three gapless points are the transition points: $V^\prime/\lambda=2.635(5), 2.92(1)$ and $3.285(10)$. 
The four phases from left to right are a MI phase, a yFM phase, a SPT phase and a zFM phase.
$\Delta_b$ is the bulk gap in the SPT phase.}
\label{FIG3}
\end{figure}
We found that these neutral excitation gaps behave significantly different in different phases.
To illustrate this, we plot the gaps $\Delta_1$ and $\Delta_2$ as a function of $V^\prime/\lambda$ for $t/\lambda=0.4$ in Fig. \ref{FIG3}. 
We clearly see four gapped phases, which are separated by gapless transition points. These transition points are 
well consistent with those determined by the order parameters.
In the MI phase, both $\Delta_1$ and $\Delta_2$ are finite. Moreover, $\Delta_1 = \Delta_2$. 
The ground state is unique and no gapless edge modes are found in this phase. 
In yFM and zFM phases, time-reversal symmetry is spontaneously broken. 
This is firmly reflected in the neutral excitation gaps. In both phases, $\Delta_1$ is zero but $\Delta_2$ is finite. 
This indicates that the ground states are twofold degenerate. 
In the SPT phase, both $\Delta_1$ and $\Delta_2$ are zero. These gapless excitations are confirmed to be edge modes. 
To extract the bulk excitation gap in the SPT phase with OBC, we lift the degeneracy by adding 
a chemical potential $\mathrm{\mu_{edge}}$ but with opposite sign to the two ends of the chain \cite{KUHNER1,TORRE1}. By properly adjusting 
the $\mathrm{\mu_{edge}}$, we can obtain\cite{SUPP1} the bulk excitation gap $\Delta_b$.
We confirmed that the gap obtained by this method is equivalent to that obtained with periodic boundary condition (PBC). 
Moreover, the ground state is unique with PBC. These are the hallmark of an SPT phase \cite{XUCHENKE1}. 

\begin{figure}
\includegraphics[width=7.7cm, clip]{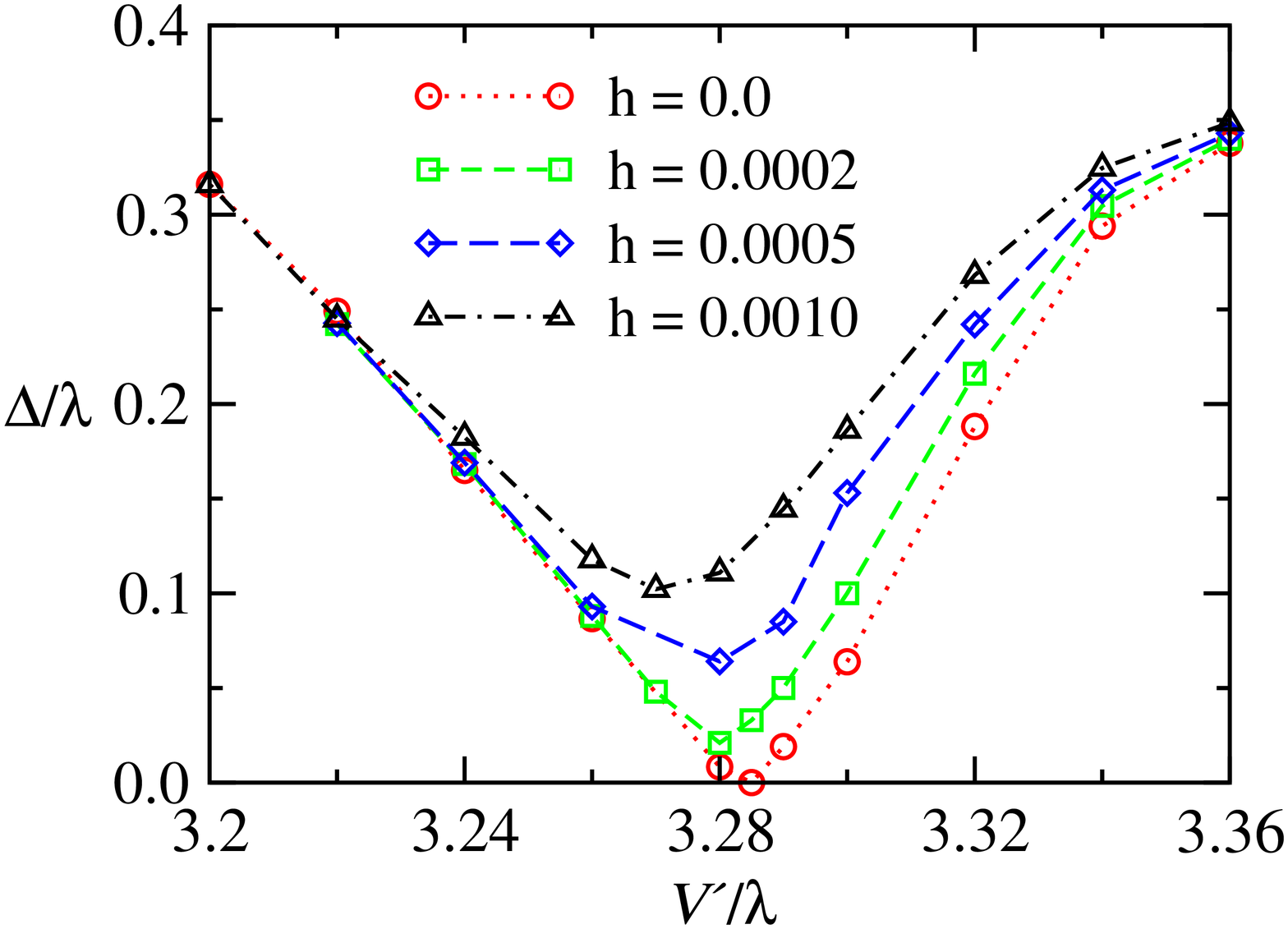}
\caption{(color online) Neutral excitation gaps as a function of $V^\prime$ for various magnetic field at $t/\lambda=0.4$.
$\Delta$ represents $\Delta_2$ for $h=0.0$ and $\Delta_1$ otherwise.
At $h=0.0$, the gap closes at the transition point from SPT phase to zFM phase. But for finite $h$, the two phases can be
adiabatically connected without gap closure.
A chemical potential $\mathrm{\mu_{edge}}=\pm{4}$ is added respectively to the two ends of the chain to lift the edge degeneracy.}
\label{FIG4}
\end{figure}
A key understanding of the SPT phase is to find out the symmetry that stabilizes the SPT phase. 
In the extended Bose-Hubbard model, the SPT phase is protected by inversion symmetry. 
However, inversion symmetry is explicitly broken in the presence of the SOC.
In our model (\ref{HSOC}), we will demonstrate that the SPT phase is protected by time-reversal symmetry. 
To show this, we perturbate the Hamiltonian (\ref{HSOC}) by a Zeeman term $h\sum_iS_i^z$, which breaks time-reversal symmetry. 
In Fig. \ref{FIG4}, we plot the neutral excitation gaps as a function of $V^\prime/\lambda$ near the transition point from 
the SPT phase to the zFM phase for various $h$. 
When $h=0$, a gap closure occurs at the transition point.
However, once $h$ becomes finite but still very small, we do not observe such gap closure. 
In this case, the SPT phase and the zFM phase can adiabatically evolve into each other. 
This provides a direct evidence that the SPT phase is protected by time-reversal symmetry.
Therefore, according to the group cohomology theory, this SPT phase is the Haldane phase \cite{CHEN2}.

Haldane phase is usually identified by a nonlocal string order \cite{TORRE1, BERG1}, 
which can now be observed \cite{ENDRES1} in quantum gas. However, in our model, this phase is 
protected by time-reversal symmetry. There is no conventional string order \cite{POLLMANN1}. 
\begin{figure}
\includegraphics[width=7.7cm, clip]{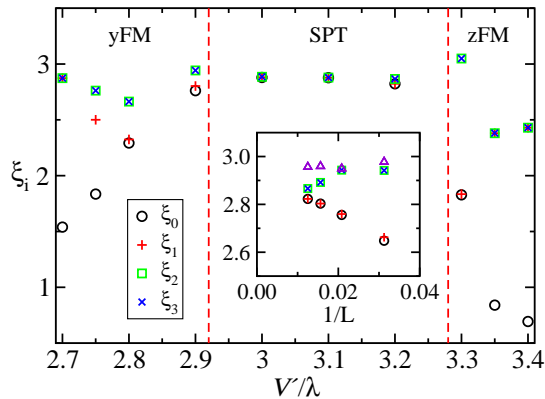}
\caption{(color online) Four lowest entanglement spectra for $L=80$ as a function of $V^\prime/\lambda$ at $t/\lambda=0.4$.
The entanglement spectra are calculated by DMRG with periodic boundary conditions and 2000 states kept.
The accidental degeneracy in the symmetry-broken phases results from the cat states \cite{POLLMANN2}. 
Inset: finite-size extrapolation of the eight lowest entanglement spectra for $V^\prime/\lambda=3.2$ as 1/L. 
Four entanglement spectra represented by triangles ($\bm{\textcolor{violet}{\triangle}}$) almost overlap.  
}
\label{FIG5}
\end{figure}
Instead, Haldane phase is generally characterized by the
even-fold degeneracy \cite{POLLMANN1} of the entanglement spectrum \cite{LI1}.
This can be verified in our DMRG simulations.
The entanglement spectra are defined as $\xi_i=-\ln(\rho_i)$ with $\rho_i$ the eigenvalues of the reduced 
density-matrix, which is readily available in our DMRG simulations. In our work, the reduced-density matrix 
is obtained by tracing out half of a chain.
In the main panel of Fig. \ref{FIG5}, we present four lowest entanglement spectra at $t/\lambda=0.4$ for $L=80$. In the deep Haldane 
phase, the four entanglement spectra are already degenerate (within our numerical error). 
In the vicinity of the right boundary, slight splitting of the entanglement spectrum is observed. 
This is just a finite-size effect. To show this, 
in the inset of Fig. \ref{FIG5}, we plot the lowest eight entanglement spectra for $V^\prime/\lambda=3.2$ 
as a function of $1/L$. The lowest four spectra merge as the size increases and 
the other four already overlap for our finite sizes.  
We also check some other spectrum and confirm the degeneracy is a multiple of four. This provides robust numerical evidence for the 
SPT phase. We want to mention that the degeneracy of the entanglement spectrum is closely related to the boundary conditions. 
With OBC, the characteristic degeneracy will become twofold\cite{SUPP1}.

In conclusion, by using density-matrix renormalization group method, we investigate numerically a one-dimensional 
correlated bosonic lattice model with a synthetic SOC that was realized in recent experiments. 
We found an SPT phase in the phase diagram. Such an SPT 
phase has a unique ground state, gapful bulk excitations for PBC as well as gapless edge modes for OBC. 
In contrast to the extended Bose-Hubbard model, this model has no inversion symmetry due to 
the presence of the spin-orbit coupling. We demonstrate numerically that the SPT phase is protected 
by time-reversal symmetry. It is then classified into a Haldane phase. We confirm this by the degeneracy of the entanglement spectrum.
In addition to this topologically nontrivial phase, we find four other phases in the phase diagram:
a Mott insulating phase with no long range order, a yFM superfluid and a zFM insulating phase, where time-reversal
symmetry is spontaneously broken, and a density-wave phase, where the translational symmetry is spontaneously broken.
The SPT phase can be identified by measuring the dynamic structure factors \cite{EJIMA1, BERG1}.
The propagation of gapless edge modes in the SPT phase can be directly observed in ultracold systems \cite{GOLDMAN1}.
The magnetic phases and DW phase are detectable by the spin-dependent Bragg scattering \cite{CORCOVILOS1} and by
quantum noise corrections \cite{ALTMAN1}, respectively. 

We thank Frank Pollmann, Masaki Oshikawa, Zheng-Xin Liu, Andreas Schulz, Sebastian Eggert, 
Yi-Fei Wang, Peng Zhang, and Xuefeng Zhang for helpful discussions. 
This work was supported by the National Natural Science Foundation of China (Grants No. 11474029, No. 91321103), 
by the National Basic Research Program of China (973 program) (Grant No. 2015CB921103), 
and by the SFB Transregio 49 of the Deutsche Forschungsgemeinschaft (DFG) and the Allianz f\"ur Hochleistungsrechnen Rheinland-Pfalz (AHRP).
The computational resources were provided partly by the high-performance computer-Kohn at the Physics Department, RUC.

\end{bibunit}

\begin{bibunit}

\clearpage
\pagebreak
\onecolumngrid
\widetext
\begin{center}
{\textbf{Supplemental Material for: \\ \textit{Symmetry-protected topological phase in a one-dimensional correlated bosonic model with a synthetic spin-orbit coupling}}}
\end{center}

\setcounter{equation}{0}
\setcounter{figure}{0}
\setcounter{table}{0}

In this Supplemental Material we present the details on the transformation of the interaction, some numerical techniques,
the phase transitions in our phase diagram and the excitation gaps.

\section{Transformation of the interaction}

In one dimension, the SOC can be eliminated by the local gauge transformation \cite{S-ZHAO1}
\begin{eqnarray}
\left(\begin{array}{c}
\hat{c}_{i\uparrow}\\
\hat{c}_{i\downarrow}
\end{array}\right) & = & \left(\begin{array}{cc}
\cos\frac{\omega_i}{2} & -\sin\frac{\omega_i}{2}\\
\sin\frac{\omega_i}{2} & \cos\frac{\omega_i}{2}
\end{array}\right)\left(\begin{array}{c}
\hat{c}_{i\uparrow}^{\prime}\\
\hat{c}_{i\downarrow}^{\prime}
\end{array}\right).
\label{S-EROTATE}
\end{eqnarray}
with the requirement that $\omega_{i+1}-\omega_i = 2\arctan(\frac{\lambda}{t})$.
Under this transformation, the particle number $\hat{n}_i\left(=\hat{n}_{i\uparrow}+\hat{n}_{i\downarrow}\right)$ is also invariant.
Therefore, if the interaction is in this form $\mathcal{H}_{int}=\sum_{ij}A_{ij}\hat{n}_{i}\hat{n}_{j}$,
it is invariant by the transformation (\ref{S-EROTATE}). However, for general case, the interaction is not invariant
under such a transformation. To see this, let us just consider the onsite
interaction at site $i$,
$\frac{U}{2}\sum_{\tau}\hat{n}_{i\tau}\left(\hat{n}_{i\tau}-1\right)+U^\prime{\hat{n}_{i\uparrow}\hat{n}_{i\downarrow}}$.
It can be written as
$\frac{U}{2}\hat{n}_i\left(\hat{n}_i-1\right)+\left(U^\prime-U\right)\hat{n}_{i\uparrow}\hat{n}_{i\downarrow}$.
Obviously, under the transformation (\ref{S-EROTATE}), the first term is always invariant but
the second term is not invariant if $U\ne{U^\prime}$.

\section{Infinite-DMRG algorithm for the entanglement spectrum and central charge}
The ground state of a one-dimensional quantum system (with local bases $|s_{i}\rangle$ for the $i$-th site)
in the thermodynamical limit can be represented by a
matrix-product state (MPS)(like the infinite-DMRG (iDMRG)) \cite{McCulloch}
\begin{equation}
|\psi\rangle = \sum_{\{s_{i}\}} \cdots A\left(s_{i-1}\right)  A\left(s_{i}\right) \Lambda B\left(s_{i+1}\right) B\left(s_{i+2}\right) \cdots |s_{i-1}\rangle |s_{i}\rangle |s_{i+1}\rangle |s_{i+2}\rangle\cdots
\end{equation}
where $\sum_{s_{i}} A^{\dagger} \left(s_{i}\right) A\left(s_{i}\right)=\mathbb{I}$ and $\sum_{s_{i}} B \left(s_{i}\right) B^{\dagger}\left(s_{i}\right)=\mathbb{I}$
and $\Lambda$ (on the center bond) contains singular values.
In the thermodynamical limit, any bond can be a center and thus $A\Lambda=\Lambda B$. In a period of two sites the wave function reads
\begin{equation}
|\psi\rangle = \sum_{\{s_{i}\}} \cdots \left.\Lambda B\left(s_{i-1}\right) \Lambda^{-1}\right) \left(A\left(s_{i}\right) \Lambda B\left(s_{i+1}\right) \Lambda^{-1}\right) \left(A\left(s_{i+2}\right)\Lambda \right.\cdots |s_{i-1}\rangle |s_{i}\rangle |s_{i+1}\rangle |s_{i+2}\rangle\cdots
\end{equation}
In general, a correlation function can be wirtten as
\begin{equation}
\left\langle {\hat O}_{0} {\hat O}_{2r+1} \right\rangle = \rm{Tr}\left({\mathbb T}_{h} \left({\mathbb T}_{A} {\mathbb T}_{B}\right)^{r} {\mathbb T}_{e} \right) = \rm{Tr}\left({\mathbb T}_{h} {\mathbb T}^{r} {\mathbb T}_{e} \right)
\end{equation}
where ${\mathbb T}_{h}=\sum_{s, s'} {\hat O}_{s', s} A^{*}\left(s'\right) \otimes A\left(s\right)$,
${\mathbb T}_{e}=\sum_{s, s'} {\hat O}_{s', s} \left(\Lambda B^{*}\left(s'\right)) \otimes (\Lambda B\left(s\right)\right)$
and the transfer matrix ${\mathbb T}_{A}=\sum_{s} \left(A^{*}\left(s\right)\Lambda\right) \otimes \left(A\left(s\right)\Lambda\right)$ and
${\mathbb T}_{B}=\sum_{s} (B^{*}\left(s\right)\Lambda^{-1}) \otimes (B\left(s\right)\Lambda^{-1})$ for the site A and B, respectively.
We suppose the mixed transfer matrix ${\mathbb T} = {\mathbb T}_{A} {\mathbb T}_{B} = \sum_{l} | R_{l}\rangle \alpha_{l} \langle L_{l}|$
with the left eigenvector $| L_{l}\rangle$ and the right eigenvector $|R_{l}\rangle$ for the $l$-th largest eigenvalue $\alpha_{l}$ ($l=0$, $1$, $2$, $\cdots$).
The correlation function has a polynomial expansion
\begin{equation}
\left\langle {\hat O}_{0} {\hat O}_{2r+1} \right\rangle = \rm{Tr}\left({\mathbb T}_{h} \left[\sum_{l} | R_{l}\rangle (\alpha_{l})^{r} \langle L_{l}|\right] {\mathbb T}_{e} \right).
\end{equation}
For sufficiently large $r$, $\left\langle {\hat O}_{0} {\hat O}_{2r+1} \right\rangle \approx G_{0} + G_{1}$
where the $0$-th order term $G_{0}$ vanishes after normal-ordering if there is no long-range order,
the $1$-st order term $G_{1} \propto \exp(-(2r+1)/\zeta)$ with the correlation length $\zeta=2/\ln(\alpha_{0}/\alpha_{1})$.

The entanglement between the left and right semi-infinite chain is usually measured by
the von Neumann entanglement entropy $S=\rm{Tr}\left(2\Lambda^{2}\ln\Lambda\right)$. The entanglement spectrum is therefore defined as
$\xi_{i}=-2\ln\Lambda_{i}$.
In a one-dimensional gapped quantum system, the entanglement entropy is finite \cite{Arealaw}.
But at the critical point, the entanglement entropy grows logarithmically with the chain length.
However, due to the finite truncation dimension of the $A$ and $B$ matrices, both the $S$ and
$\zeta$ actually remain finite, which satisfy \cite{Finiteentangle}
\begin{equation}
S \sim \frac{c}{6} \ln\zeta,
\end{equation}
where $c$ is the central charge \cite{S-CARDY1}. On the condition that we can locate the critical point accurately,
the central charge is then determined from this linear relation \cite{Shijie}.

\section{Central charges and Criticality}
For a one-dimensional finite system with length $L$ and periodic boundary condition (PBC),
conformal field theory predicts \cite{S-CARDY1} that
the von Neumann  entanglement entropy $S_L(l)$ between the subsystem of length $l$ and the remaining part is given by
$S_L(l)=\frac{c}{3}\ln[\frac{l}{\pi}\sin(\frac{\pi{l}}{L})]+s_1$, where $c$ is the central charge and
$s_1$ is a nonuniversal constant. An accurate estimator of the central
charge in DMRG is then given by \cite{S-NISHIMOTO1}
\begin{equation}
c^\star(L)=\frac{3[S_L(\frac{L}{2}-1)-S_L(\frac{L}{2})]}{\ln[\cos(\frac{\pi}{L})]}.
\label{CC}
\end{equation}
\begin{figure}[h]
\includegraphics[width=8.4cm, clip]{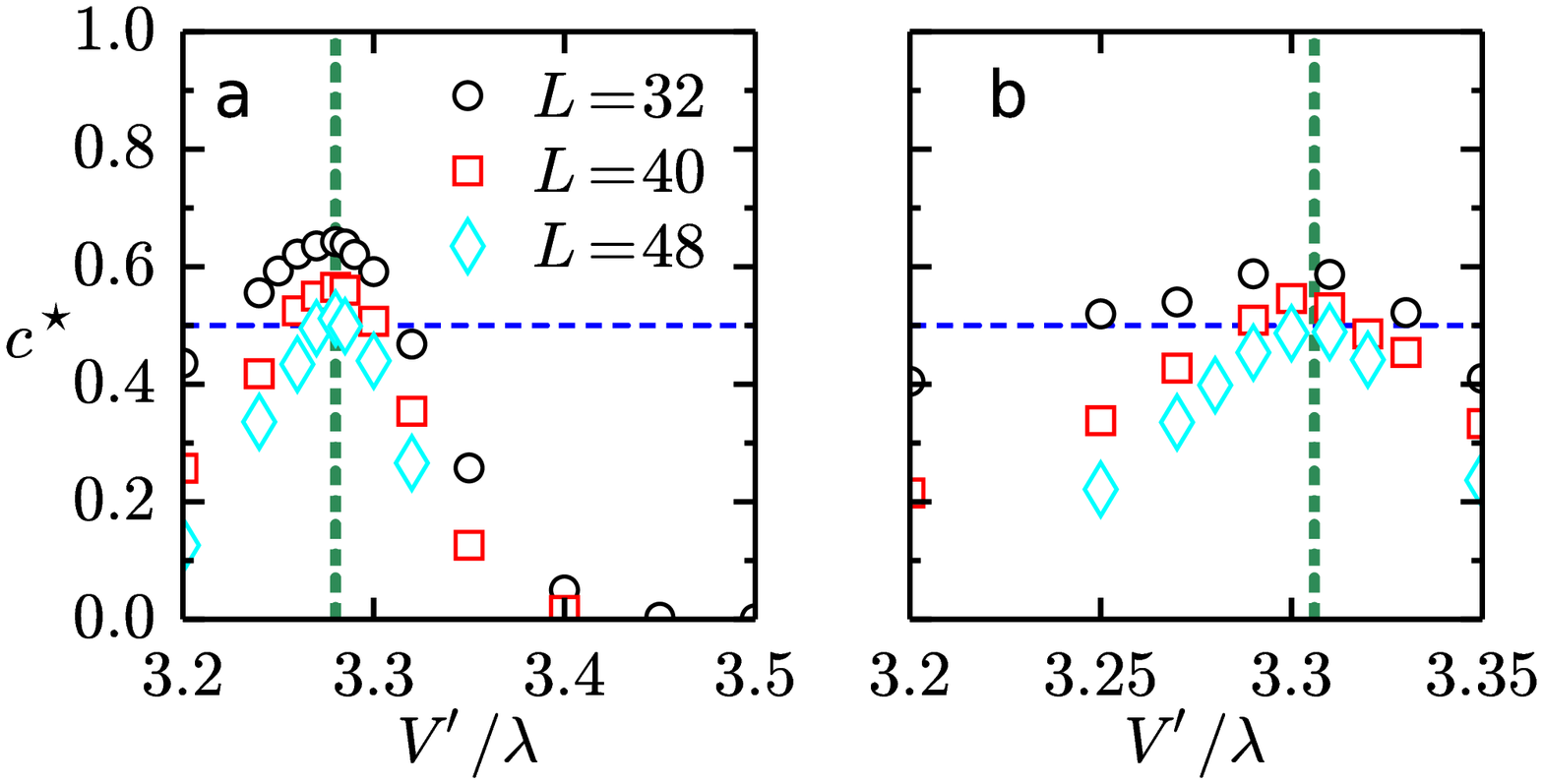}
\vspace{-0.3cm}
\caption{(color online) The central charge $c^\star$ estimated from the von Neumann entropy
for finite-size systems near (a) the SPT-zFM transition at $t/\lambda=0.4$ and (b) the SPT-DW transition at $t/\lambda=0.6$.
The estimated transition points are marked by the vertical dashed lines.
At both transition points, $c\simeq{0.5}$, suggesting an Ising universality class.}
\label{S-FIG1}
\end{figure}
This method has been successfully used to estimate the central charges at critical points in a variety of bosonic models.
In order to get accurate results, we add two components (instead two sites) in each DMRG iteration.
We keep at most 5000 states. Sweeps are performed to improve the accuracy after the required length is reached.
In Fig. \ref{S-FIG1}, we see that at both the SPT-zFM and SPT-DW transition points the central charges show a peak.
These critical points are 3.280(5) and 3.306(5), which agree well with those estimated from the order parameters and excitation gaps.
And away from the critical point, $c$ falls off rapidly as $L$ increases, thus suggesting a fully gapped phase.
At both the critical points, $c$ are estimated to be about $0.5$, and thus both transitions belong to Ising universality class.
These conclusions are consistent with the fact that the ground states are twofold degenerate in the zFM phase,
and the $Z_2$ symmetry is spontaneously broken in the DW phase.

At those transition points with $c\ge{1}$, we find that it is extremely difficult to extract
reliably the central charges with PBC due to the finite-size effect. Instead, we resort to iDMRG to estimate the central charges.
\begin{figure}[h]
\includegraphics[width=16.8cm, clip]{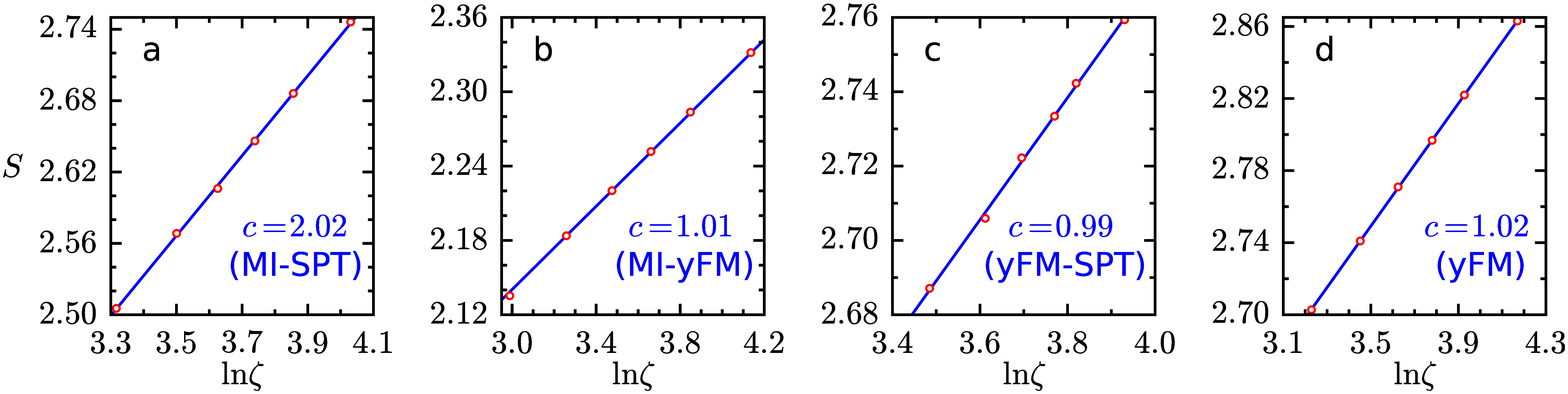}
\vspace{-0.3cm}
\caption{(color online) Entanglement entropy $S$ are shown as a function of $\ln\zeta$ at the critical points and in yFM phase.
(a) For $t/\lambda=0.0$, the critical point for MI-SPT transition is $V^{\prime}/\lambda=2.763(1)$.
For $t/\lambda=0.4$, the critical points for MI-yFM and yFM-SPT transition
are (b) $V^{\prime}/\lambda=2.6282(5)$ and (c) $V^\prime/\lambda=2.921(2)$,
respectively. (d) The central charge for yFM phase is calculated at $V^{\prime}/\lambda=2.8$, $t/\lambda=0.4$.}
\label{S-FIG2}
\end{figure}
We want to stress that yFM deserves particular attention.
In this phase, due to the presence of the cat state, the dominant eigenvalues of the transfer matrix are doubly degenerate.
In this case, the third eigenvalue is actually the second largest one. Therefore, the definition of the
correlation length is modified correspondingly.
In Fig. \ref{S-FIG2}, we illustrate the central charges determined by iDMRG.
When $t/\lambda=0$, $c$ is determined to be 2 at the MI-SPT transition point, as shown in panel a.
This value can be understood as follows.
Recall that we can interchange $t$ and $\lambda$ as well as $V$ and $V^\prime$ \cite{S-ZHAO1}.
When $t/\lambda=0$, the model has $U(1)\times{U(1)}$ symmetry. Because both the MI and SPT phase are gapful,
such a model is adiabatically connected to two decoupled extended Bose-Hubbard model. Thus, at the critical point,
it is described by a two-component Tomonaga-Luttinger liquid. When $t/\lambda$ is finite, the $U(1)\times{U(1)}$ symmetry
is reduced to $U(1)$, i.e., only the total particle number is conserved. This is reflected by the central charge $c=1$
in yFM phase. Later we will show that it is a superfluid phase.
We thus expect that it behaves as a one-component Tomonaga-Luttinger liquid.
The central charges are estimated to be 1 at both the MI-yFM and yFM-SPT transition points.
Such transitions are expected to be the same as those in the extended Bose-Hubbard model, i.e., they are the XY type.

In the zFM and DW phase(yFM and DW phase as well), different symmetries are spontaneously broken. Therefore, we expect that both the
zFM-DW and yFM-DW transitions are of the first order.
\begin{figure}[h]
\includegraphics[width=8.4cm, clip]{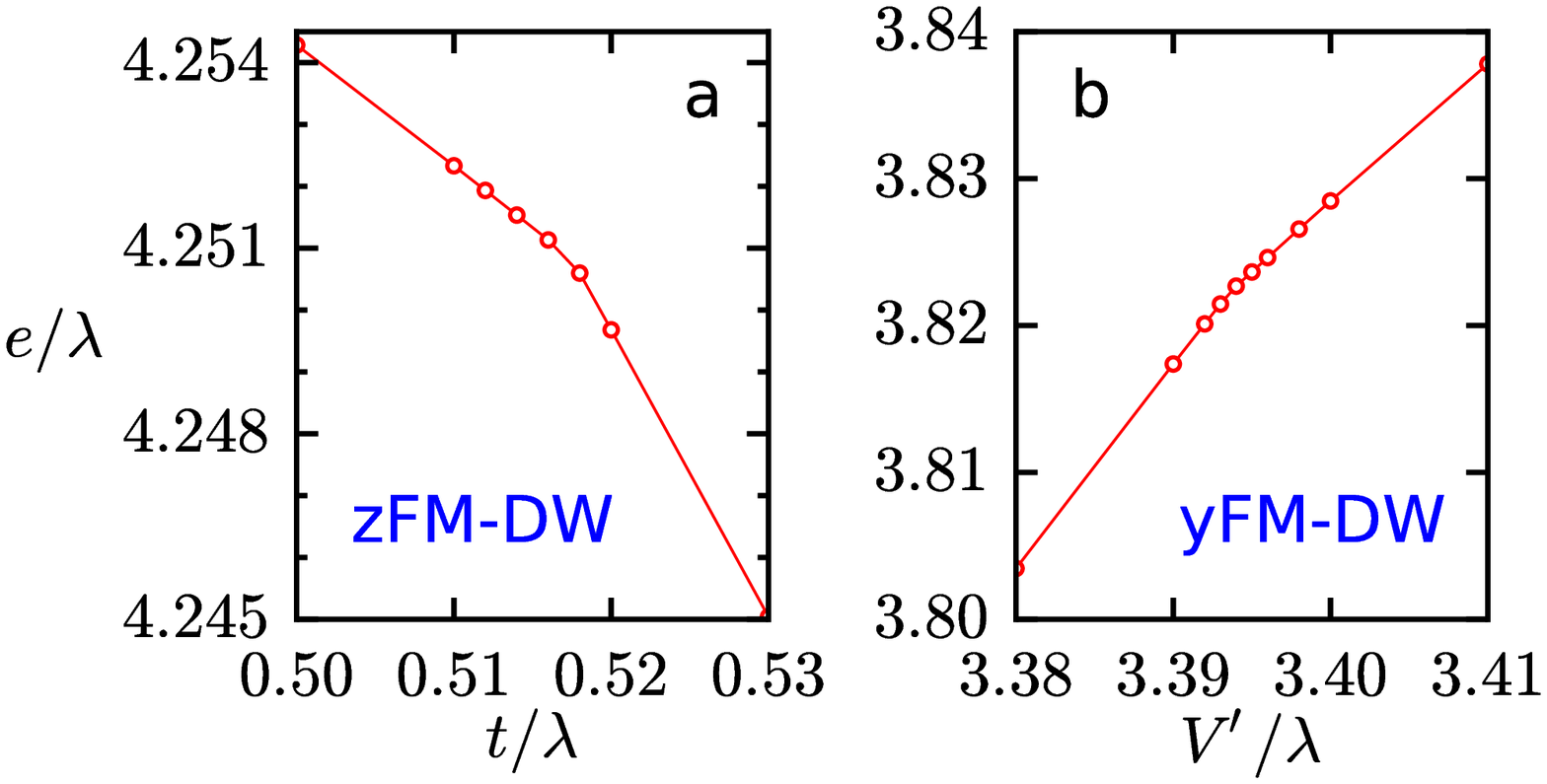}
\vspace{-0.1cm}
\caption{(color online) Energy per site $e$ (obtained by iDMRG with $m=1280$) are shown as (a) a function of
$t$ with $V^{\prime}/\lambda=3.5$ for the zFM-DW transition and (b) as a function of $V^\prime$ with
$t/\lambda=1.0$ for the yFM-DW transition.
The kinks of the energy unambiguously suggest first-order phase transitions. The transition points are
$t/\lambda=0.5175(2)$ and $V^{\prime}/\lambda=3.3936(2)$, respectively.}
\label{S-FIG3}
\end{figure}
As illustrated in Fig. \ref{S-FIG3}, a first-order phase transition is signaled by a kink of the ground state energy.
The transition points locate at $t/\lambda=0.5175(2)$ for $V^{\prime}/\lambda=3.5$
for the zFM-DW transition and at $V^{\prime}/\lambda=3.3936(2)$ for $t/\lambda=1.0$ for
the yFM-DW transition. These transition points agree well with those obtained from the excitation gaps in our main text.

\section{Entanglement spectrum}
\begin{figure}[h]
\includegraphics[width=8.4cm, clip]{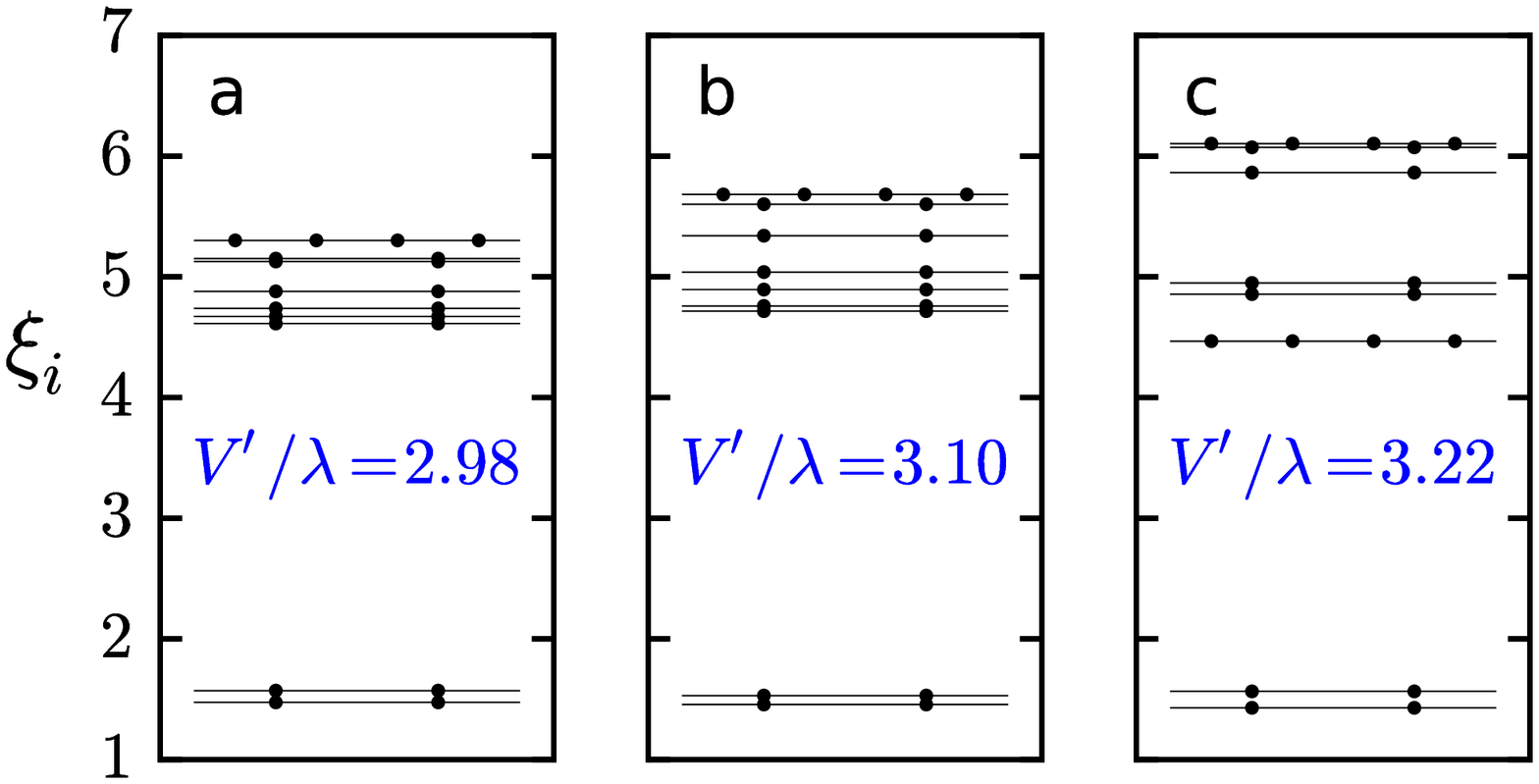}
\vspace{-0.1cm}
\caption{(color online) Entanglement spectra in the SPT phase obtained by iDMRG are shown for
$V^{\prime}/\lambda=2.98$, $3.10$ and $3.22$ with $t/\lambda=0.4$. The degeneracy of all levels is a multiple of two,
which is the characteristic feature of the SPT phase.}
\label{S-FIG4}
\end{figure}

In the main text, the entanglement spectrum obtained by DMRG with PBC are a multiple of four in SPT phase.
This is because there are two cuts with PBC. With one cut, the entanglement spectrum is expected to be a multiple of 2.
To confirm this, we perform iDMRG calculations and our results are shown in Fig. \ref{S-FIG4}.
As anticipated, all the entanglement spectrum are at least twofold degenerate \cite{frank} in the SPT phase.

\section{Lift the degeneracy by edge chemical potential}
The edge degeneracy in the symmetry-protected topological phase (SPT) can be lifted by adding
a chemical potential $\mathrm{\mu_{edge}}$ to the left end and $\mathrm{-\mu_{edge}}$ to
the right end of the chain \cite{S-KUHNER1, S-TORRE1}.
\begin{figure}[h]
\includegraphics[width=8.4cm, clip]{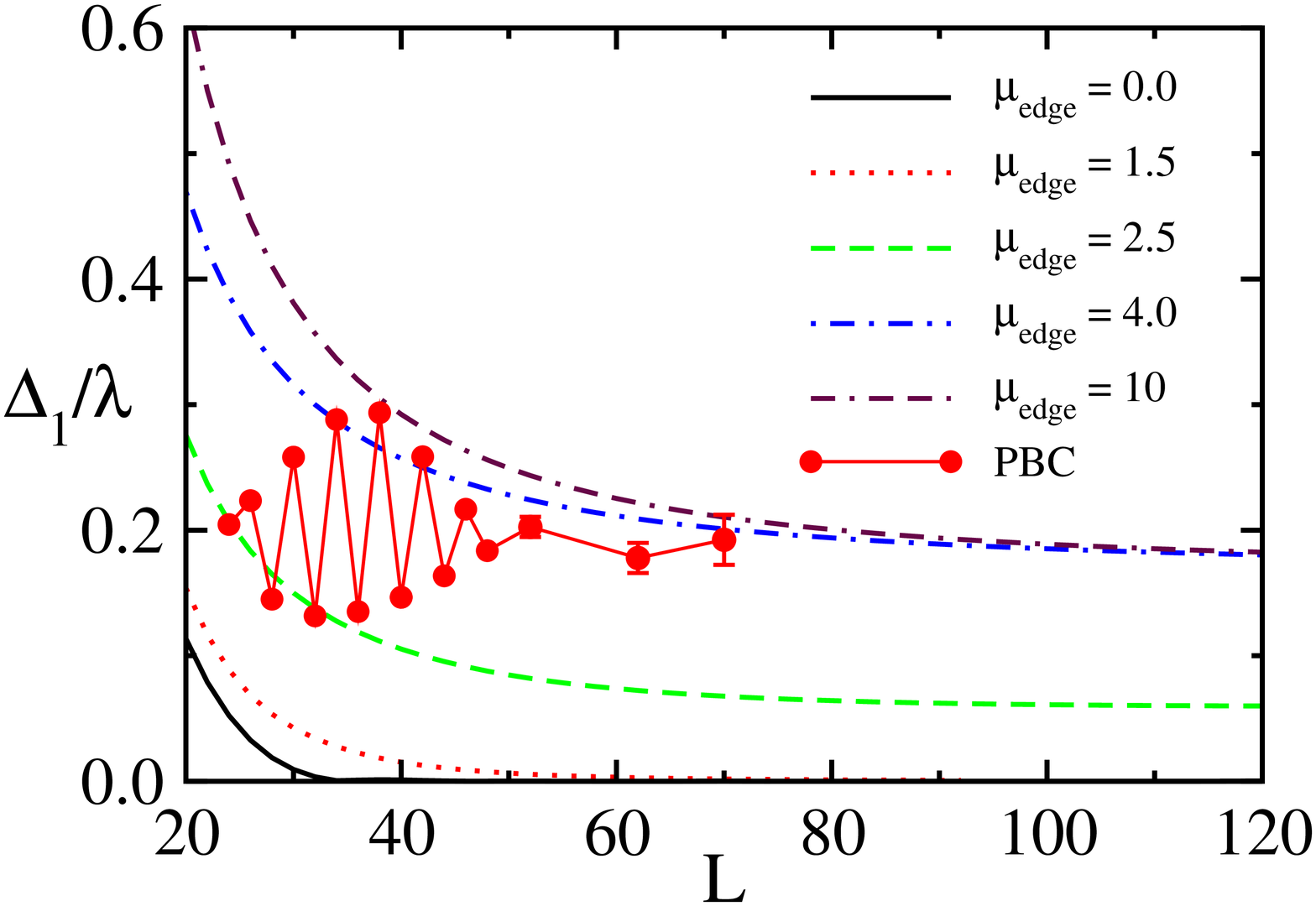}
\vspace{-0.3cm}
\caption{(color online) The bulk excitation gap is extracted with open boundary condition by adding edge chemical
potential $\pm\mathrm{\mu_{edge}}$ to the two ends of the chain,
respectively. When the $\mathrm{\mu_{edge}}$ is sufficiently large, a finite gap appears
in the spectrum. This gap becomes independent of $\mathrm{\mu_{edge}}$ in the thermodynamic limit
after $\mathrm{\mu_{edge}}$ is larger than a critical value. We confirmed that the gap we obtain is really a bulk gap by performing
DMRG calculation with the PBC, as we show in red solid circles.}
\label{S-FIG5}
\end{figure}
Fig. \ref{S-FIG5} illustrates this procedure for $V^\prime/\lambda=3$, $t/\lambda=0.4$.
When $\mathrm{\mu_{edge}}=1.5$, the edge modes remain gapless. However, the edge modes
become gapful for $\mathrm{\mu_{edge}} = 2.5$. This excitation gap increases and approaches a constant,
marked as $\Delta_b$, as $\mathrm{\mu_{edge}}$ increases. $\mathrm{\mu_{edge}}$ increases further does not change $\Delta_b$, as the data for
$\mathrm{\mu_{edge}} = 4$ and $\mathrm{\mu_{edge}}=10$ show. This $\Delta_b$ turns out to be the bulk excitation gap.
To confirm this, we performed DMRG calculations with PBC. In this calculation, sweeps are performed to
improve the accuracy for L = 62 and 72. This figure demonstrates that the gap is equivalent to $\Delta_b$ within our error bar.

\section{Charge gap}
An insulating phase and a superfluid phase are characterized by a gapful and
gapless charge excitation, respectively. Such charge gap is defined as
\begin{equation} 
\Delta_c=E_0(N+1,L)+E_0(N-1,L)-2E_0(N,L)
\end{equation}
for our model with length $L$ and boson number $N$. Here $N=2L$. Due to the edge excitations in the SPT phase,
the charge gap may be obtained by DMRG with the PBC or with open boundary condition
in the presence of sufficiently large edge chemical potentials, as we describe in last section.
\begin{figure}
\includegraphics[width=8.4cm, clip]{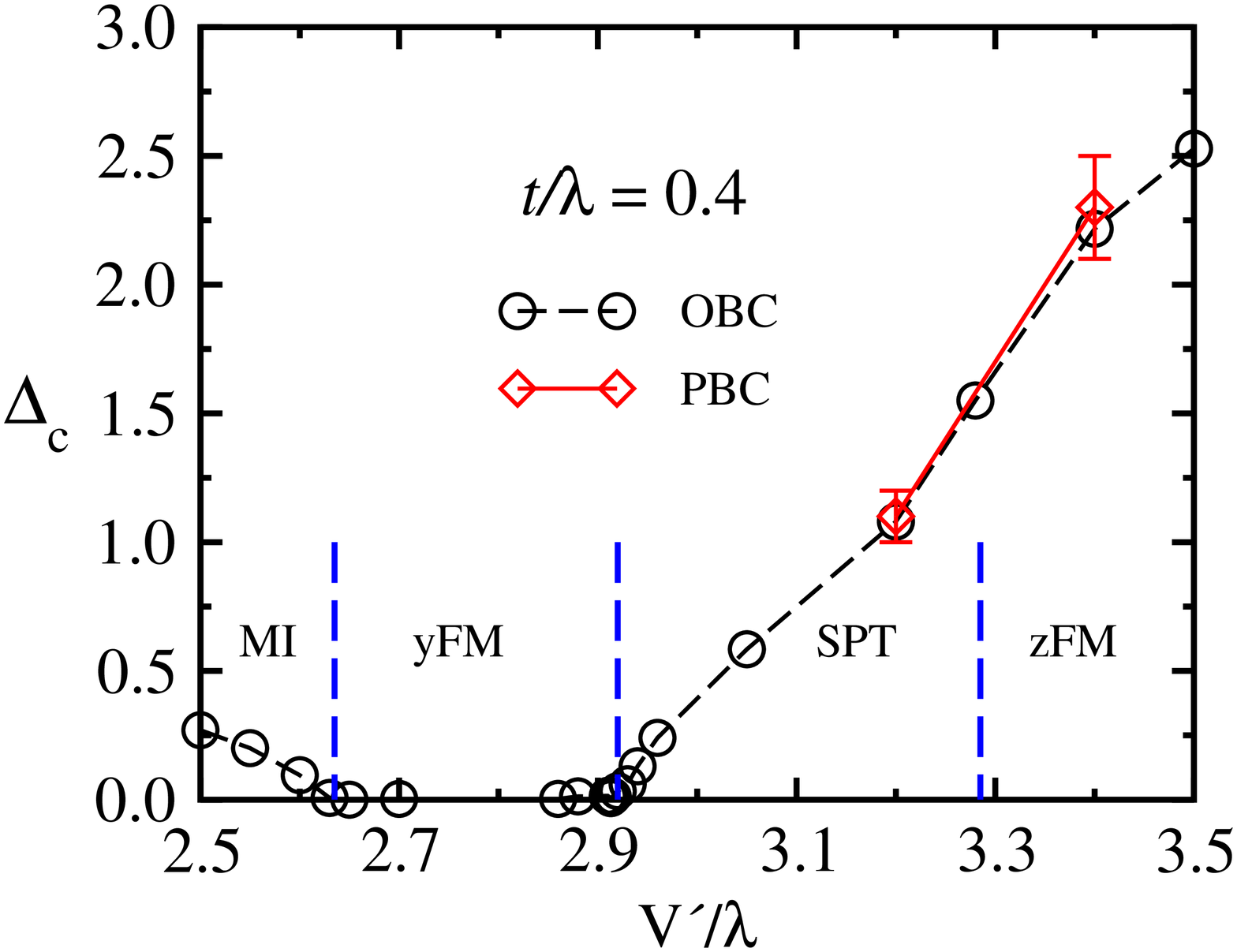}
\caption{(color online) Charge gaps as a function of $V^\prime/\lambda$ for $t/\lambda=0.4$. Data represented by open circle $\bigcirc$
are calculated with open boundary condition. Data obtained with PBC ($\Diamond$) are shown for comparison.
Vertical blue dashed lines mark the critical points obtained from neutral excitation gaps.}
\label{CHARGEGAP}
\end{figure}
In Fig. \ref{CHARGEGAP}, we show the charge gaps as a function of $V^\prime/\lambda$ for $t/\lambda=0.4$. The other parameters are
the same as those in Fig. 1 in the main text. In the yFM phase, $\Delta_c$ is zero and we conclude that it is a superfluid phase.
In other phases $\Delta_c$ is finite($\Delta_c$ in DW phase is also finite but not shown here). Therefore, they are insulating phases.
These results are consistent with our data of the central charges.

\end{bibunit}


\begin{thebibliography}{99}
\bibitem{GU1} Z.-C. Gu and X.-G Wen, Phys. Rev. B {\bf{80}}, 155131 (2009).
\bibitem{POLLMANN1} F. Pollmann, A. M. Turner, E. Berg, and M. Oshikawa, Phys. Rev. B {\bf{81}}, 064439 (2010).
\bibitem{SCHUCH1} N. Schuch, D. Pe\'rez-Garcia, and I. Cirac, Phys. Rev. B {\bf{84}}, 165139 (2011).
\bibitem{CHEN1} X. Chen, Z.-C. Gu, and X.-G. Wen, Phys. Rev. B {\bf{84}}, 235128 (2011).
\bibitem{CHEN2} X. Chen, Z.-C. Gu, Z.-X. Liu, and X.-G. Wen, Phys. Rev. B {\bf{87}}, 155114 (2013).
\bibitem{HASAN1} M.Z. Hasan and C.L. Kane, Rev. Mod. Phys. {\bf{82}}, 3045 (2010).
\bibitem{QI1} X.-L. Qi and S.-C. Zhang, Rev. Mod. Phys. {\bf{83}}, 1057 (2011).
\bibitem{KANE1} C. L. Kane and E. J. Mele, Phys. Rev. Lett. {\bf{95}}, 226801 (2005).
\bibitem{LIN1} Y. J. Lin, K. Jim\'enez-Garc\'ia, and I. B. Spielman, Nature {\bf{471}}, 83 (2011).
\bibitem{WANG1} P. J. Wang, Z. Q. Yu, Z. K. Fu, J. Miao, L. H. Huang, S. J. Chai, H. Zhai, and J. Zhang, Phys. Rev. Lett. {\bf{109}}, 095301 (2012).
\bibitem{CHEUK1} L. W. Cheuk, A. T. Sommer, Z. Hadzibabic, T. Yefsah, W. S. Bakr, and M. W. Zwierlein, Phys. Rev. Lett. {\bf{109}}, 095302 (2012).
\bibitem{ZHANG1}J.-Y. Zhang, S.-C. Ji, Z. Chen, L. Zhang, Z.-D. Du, B. Yan, G.-S. Pan, B. Zhao, Y.-J. Deng, H. Zhai, S. Chen, and J.-W. Pan, Phys. Rev. Lett. {\bf{109}}, 115301 (2012).
\bibitem{WU1} C. Wu , I. M. Shem, and X. Zhou, Chin. Phys. Lett. {\bf{28}}, 097102 (2011); X. Zhou, Y. Li, Z. Cai, C. Wu, J. Phys. B: At. Mol. Opt. Phys. {\bf{46}} 134001 (2013).
\bibitem{BIJL1} E. van der Bijl and R. A. Duine, Phys. Rev. Lett. {\bf{107}}, 195302 (2011).
\bibitem{ISKIN1} M. Iskin and A. L. Subasi, Phys. Rev. {\bf{87}}, 063627 (2013).
\bibitem{ZHAI1} H. Zhai, Int. J. Mod. Phys. B {\bf{26}} 1230001 (2012).
\bibitem{COLE1} W. S. Cole, S. Z. Zhang, A. Paramekanti, and N. Trivedi, Phys. Rev. Lett. {\bf{109}}, 085302 (2012).
\bibitem{HU1} H. Hu, B. Ramachandhran, H. Pu, and X. J. Liu, Phys. Rev. Lett. {\bf{108}}, 010402 (2012).
\bibitem{HAN1} W. Han, G. Juzeli\"unas, W. Zhang, and W.-M Liu, Phys. Rev. A {\bf{91}}, 013607 (2015).  
\bibitem{JOTZU1} G. Jotzu, M. Messer, R. Desbuquois, M. Lebrat, T. Uehlinger, D. Greif, and T. Esslinger, Nature 515, {\bf{515}}, 237 (2014).
\bibitem{BEELER1} M. C. Beeler, R. A. Williams, K. Jim\'enez-Garc\'ia, L. J. Leblanc, A. R. Perry, and I. B. Spielman, Nature {\bf{498}}, 201 (2013).
\bibitem{FU1} Z. Fu, L. Huang, Z. Meng, P. Wang, L. Zhang, S. Zhang, H. Zhai, P. Zhang, and J. Zhang, Nature physics {\bf{10}}, 110 (2014).
\bibitem{JI1} S.-C. Ji, J.-Y. Zhang, L. Zhang, Z.-D. Du, W. Zheng, Y.-J. Deng, H. Zhai, S. Chen, and J.-W. Pan, Nature physics {\bf{10}}, 314 (2014). 
\bibitem{JIMENEZ1} K. Jim\'enez-Garc\'ia, L. J. LeBlanc, R. A. Williams, M. C. Beeler, C. Qu, M. Gong, C. Zhang, and I. B. Spielman, Phys. Rev. Lett. {\bf{114}}, 125301 (2015).
\bibitem{LUO1} X. Luo, L. Wu, J. Chen, Q. Guan, K. Gao, Z. Xu, L. You, and R. Wang, arXiv:1502.07091.
\bibitem{CAI1} Z. Cai, X. Zhou, and C. Wu, Phys. Rev. A  {\bf{85}}, 061605 (2012).
\bibitem{ZHAO1} J. Zhao, S. Hu, J. Chang, P. Zhang, and X. Wang, Phys. Rev. A {\bf{89}}, 043611 (2014).
\bibitem{ZHAO2} J. Zhao, S. Hu, J. Chang, F. Zheng, P. Zhang, and X. Wang, Phys. Rev. B {\bf{90}}, 085117 (2014).
\bibitem{SUPP1} See Supplemental Material at [url] for the transformation of the interaction, the details of the numerical techniques, the discussions of the phase transitions in the phase diagram as well as the excitation gaps, which includes Refs\cite{S-EISERT1, S-POLLMANN1, S-HU1, S-NISHIMOTO1}.
\bibitem{S-EISERT1} J. Eisert, M. Cramer, and M. B. Plenio, Rev. Mod. Phys. {\bf{82}}, 277 (2010).
\bibitem{S-POLLMANN1} F. Pollmann, S. Mukerjee, A. Turner, and J. E. Moore, Phys. Rev. Lett. {\bf{102}}, 255701 (2009).
\bibitem{S-HU1} S. Hu, A. M. Turner, K. Penc, and F. Pollmann, Phys. Rev. Lett. {\bf{113}}, 027202 (2014).
\bibitem{S-NISHIMOTO1} S. Nishimoto, Phys. Rev. B {\bf{84}}, 195108 (2011).
\bibitem{HO1} T. Ho and S. Zhang, Phys. Rev. Lett. {\bf{107}}, 150403 (2011).
\bibitem{PIRAUD1} M. Piraud, Zi Cai, I. P. McCulloch, and U. Schollw\"{o}ck, Phys. Rev. A {\bf{89}}, 063618 (2014).
\bibitem{XU1} Z. Xu, W. Cole, and S. Zhang, Phys. Rev. A {\bf{89}}, 051604 (2014).
\bibitem{PEOTTA1} S. Peotta, L. Mazza, E. Vicari, M. Polini, R. Fazio, and D. Rossini, J. Stat. Mech. (2014) P09005.
\bibitem{TORRE1} E. G. Dalla Torre, E. Berg, and E. Altman, Phys. Rev. Lett. {\bf{97}}, 260401 (2006).
\bibitem{BERG1} E. Berg, E. G. Dalla Torre, T. Giamarchi, and E. Altman, Phys. Rev. B {\bf{77}}, 245119 (2008).
\bibitem{ROSSINI1} D. Rossini, and R. Fazio, New J. Phys. {\bf{14}}, 065012 (2012).
\bibitem{BATROUNI1} G. G. Batrouni, R. T. Scalettar, V. G. Rousseau, and B. Gr\'emaud, Phys. Rev. Lett. {\bf{110}}, 265303 (2013).
\bibitem{EJIMA1} S. Ejima, F. Lange, and H. Fehske, Phys. Rev. Lett. {\bf{113}}, 020401 (2014).
\bibitem{HAMNER1} C. Hamner, Yongping Zhang, M. A. Khamehchi, Matthew J. Davis, and P. Engels, Phys. Rev. Lett. {\bf{114}}, 070401 (2015).
\bibitem{DENG1} Y. Deng, J. Cheng, H. Jing, C.-P. Sun, and S. Yi, Phys. Rev. Lett. {\bf{108}}, 125301 (2012).
\bibitem{WILSON1} R. M. Wilson, B. M. Anderson, and C. W. Clark, Phys. Rev. Lett. {\bf{111}}, 185303 (2013).
\bibitem{NG1} H. T. Ng, Phys. Rev. A {\bf{90}}, 053625 (2014).
\bibitem{SYZRANOV1} S. V. Syzranov, M. L. Wall, V. Gurarie, and A. M. Rey, Nat. Comm. {\bf{5}}, 5391 (2014); M. L. Wall, K. Maeda, L. D. Carr, New J. Phys. {\bf{17}}, 025001 (2015).
\bibitem{WHITE1} S. R. White, Phys. Rev. Lett. {\bf{69}}, 2863 (1992).
\bibitem{PESCHEL1} I. Peschel, X. Q. Wang, M. Kaulke, and K. Hallberg, Density Matrix Renormalization, LNP528 (1999), Springer.
\bibitem{SCHOLLWOCK1} U. Schollw\"{o}ck, Rev. Mod. Phys. {\bf{77}}, 259 (2005).
\bibitem{MCCULLOCH1} I. P. McCulloch, arXiv: 0804.2509.
\bibitem{CALABRESE1} P. Calabrese and J. Cardy, J. Stat. Mech.: Theory Exp. (2004) P06002.
\bibitem{EJIMA2} S. Ejima, M. J. Bhaseen, M. Hohenadler, F.H.L. Essler, H. Fehske, and B. D. Simons, Phys. Rev. Lett. {\bf{106}}, 015303 (2011).
\bibitem{KUHNER1} T. D. K\"{u}hner, S. R. White, and H. Monien, Phys. Rev. B {\bf{61}}, 12474 (2000).
\bibitem{XUCHENKE1} Cenke Xu, Phys. Rev. B {\bf{87}}, 144421 (2013). 
\bibitem{ENDRES1} M. Endres, $et\ al.$, Science {\bf{334}}, 200 (2011). 
\bibitem{LI1} H. Li and F.D.M. Haldane, Phys. Rev. Lett. {\bf{101}}, 010504 (2008). 
\bibitem{POLLMANN2} F. Pollmann, A. M. Turner, Phys. Rev. B {\bf{86}}, 125441 (2012).
\bibitem{GOLDMAN1} N. Goldman, J. Dalibard, A. Dauphin, F. Gerbier, M. Lewenstein, P. Zoller, and I.B. Spielman, Proc. Natl. Acad. Sci. U.S.A. {\bf{110}}, 6736 (2013). 
\bibitem{CORCOVILOS1} T. A. Corcovilos, S. K. Baur, J. M. Hitchcock, E. J. Mueller, and R. G. Hulet, Physical Review A {\bf{81}} 013415 (2010).
\bibitem{ALTMAN1} E. Altman, E. Demler, and M. D. Lukin, Phys. Rev. A {\bf{70}}, 013603 (2004).
\end{thebibliography}

\begin{thebibliography}{99}
\bibitem{S-ZHAO1} J. Zhao, S. Hu, J. Chang, F. Zheng, P. Zhang, and X. Wang, Phys. Rev. B {\bf{90}}, 085117 (2014).
\bibitem{McCulloch} I. P. McCulloch, arXiv: 0804.2509.
\bibitem{Arealaw} J. Eisert, M. Cramer, and M. B. Plenio, Rev. Mod. Phys. {\bf{82}}, 277 (2010).
\bibitem{Finiteentangle} F. Pollmann, S. Mukerjee, A. Turner, and J. E. Moore, Phys. Rev. Lett. {\bf{102}}, 255701 (2009).
\bibitem{S-CARDY1} P. Calabrese and J. Cardy, J. Stat. Mech.: Theory Exp. (2004) P06002.
\bibitem{Shijie} S. Hu, A. M. Turner, K. Penc, and F. Pollmann, Phys. Rev. Lett. {\bf{113}}, 027202 (2014).
\bibitem{S-NISHIMOTO1} S. Nishimoto, Phys. Rev. B {\bf{84}}, 195108 (2011).
\bibitem{frank} F. Pollmann, E. Berg, A. M. Turner, and M. Oshikawa, Phys. Rev. B {\bf{81}}, 064439 (2010).
\bibitem{S-KUHNER1} T. D. K\"{u}hner, S. R. White, and H. Monien, Phys. Rev. B {\bf{61}}, 12474 (2000).
\bibitem{S-TORRE1} E. G. D. Torre, E. Berg, and E. Altman, Phys. Rev. Lett. {\bf{97}}, 260401 (2006).
\end{thebibliography}
\end{document}